# Improving Term Extraction Using Particle Swarm Optimization Techniques

Mohammad Syafrullah and Naomie Salim

**Abstract**—Term extraction is one of the layers in the ontology development process which has the task to extract all the terms contained in the input document automatically. The purpose of this process is to generate list of terms that are relevant to the domain of the input document. In the literature there are many approaches, techniques and algorithms used for term extraction. In this paper we propose a new approach using particle swarm optimization techniques in order to improve the accuracy of term extraction results. We choose five features to represent the term score. The approach has been applied to the domain of religious document. We compare our term extraction method precision with TFIDF, Weirdness, GlossaryExtraction and TermExtractor. The experimental results show that our propose approach achieve better precision than those four algorithm.

**Index Terms**—Term Extraction, Particle Swarm Optimization, Feature Selection, Text Mining.

—————————— ◆ ——————————

## 1 INTRODUCTION

Recently many experiments have been conducted for term extraction task. Literatures provide many examples of term extraction methods. Most of these are based on linguistic method, terminology and NLP method, and the others based on statistical/information retrieval method [1].

In linguistic method, most of existing approaches use shallow text processing techniques such as tokenizer, part-of-speech (POS) tagger and syntactic analyzer (parser). Text-to-Onto is one of the systems that use linguistic method that they called SMES (Saarbrucken Message Extraction System) in their system architecture to produce list of terms from the data input [2]. Another system called SVETLAN, use syntactic analyzer Sylex to find list of terms from the input text [3].

In terminology and NLP method many researchers have invented their new techniques [4], [5], [6]. In the work of [4], they use statistical measurement of frequency occurrence (C-value/ NC-value method), for the automatic extraction of multi-word terms, from English medical corpus. Park et. al [7], [8] introduced term cohesion which is used to calculate the cohesion of the multi-word terms. The measure is proportional to the co-occurrence frequency and the length of the term. Panel and Lin [5] present a language independent statistical corpus-based term extraction algorithm. In their algorithm, first, they collect bigram frequencies from a corpus and extract two-word candidates. After collecting features for each two-word candidate, they use mutual information (mi) and log likelihood ratio (LLR) to extend them to multi-word terms. All those experiment done both with English and Chinese corpora.

————————————————
- *Mohammad Syafrullah is with the Faculty of Computer Science and Information Systems, Universiti Teknologi Malaysia, 81310, Skudai, Johor, Malaysia.*
- *Naomie Salim is with the Faculty of Computer Science and Information Systems, Universiti Teknologi Malaysia, 81310, Skudai, Johor, Malaysia.*

In statistical method, statistical analysis will be performed on data gathered from the input, and this analysis will identifies the term from the data input based on the statistical rank. Most of the statistical methods for term extraction are based on information retrieval method for term indexing [9], [10]. Another works in this method can be found in the literature, such as the notion of "weirdness" in [11], domain pertinence in [12], [13] and domain specificity in [7], [8].

Terminology and NLP approach is more emphasize on the internal analysis for the term extraction within the corpus, while in the statistical method is more underline on the comparison of frequencies between domain specific and general corpora (external analysis).

## 2 RELATED WORKS

Kea is one of the extraction systems which are using statistical method. It uses TFIDF and first occurrence in the document as its features to determine the weight of each keyphrase. Kea's extraction algorithm has two stages, first is training stage (using Bayesian learning) which has the task to create a model for identifying keyphrases, using training documents. The second one is extraction stage which will choose keyphrases from a test document, using the model that has been made in the previous stage [14].

Turney [15] treats the problem of keyphrase extraction as supervised learning task. He presented two approaches to the task of learning to extract keyphrases from text. The first approach was to apply the C4.5 and the second one was using genetic algorithm. Turney's program is called Extractor. One form of this extractor is called GenEx, which is use Genitor genetic algorithm to maximize the performance (fitness) on the training process. Genitor is used to tune Extractor, but is no longer needed once the training process is complete.

Glossary Extraction [7],[8] is a glossary extraction tool that use two features which are domain specificity and term cohesion for calculating the term weight. Glossary Extraction algorithm, has the two important parts which are identification of candidate glossary items and glossary item ranking and selection. After obtaining candidate glossary items, the algorithm will rank them before selecting the final set. In the paper [7], [8], they claim that their method can improve the document-relevancy ranking compared with log likelihood ratio and mutual information.

The term extraction algorithm called Kea++ is the improvement of the original keyphrase extraction algorithm Kea. Medelyan and Witten [16] called their new approach as index term extraction, because they combine the advantages of both keyphrase extraction and term assignment into a single scheme. Their preliminary evaluations shows that the Kea++ significantly outperforms compared with Kea extraction algorithm.

Another term extraction systems called Term Extractor [12], [13], use three features to compute their term weight. Domain pertinence is used to perform a contrastive analysis between domain of interest documents and other domains documents. Domain consensus is used to measure the distribution of terms in a domain of interest, while the definition of lexical cohesion similar to that already introduced in [7], [8].

## 3 PARTICLE SWARM OPTIMIZATION

Particle swarm optimization first introduced by Kennedy and Eberhart [17], [18], [19], as an optimization technique based on the movement and intelligence of the swarm. It inspired by social behavior and dynamics of movement of birds and fish. PSO uses a number of particles that constitute a swarm moving around in the search space to find the best solution. Each particle is treated as a point in the search space which adjusts its flying according to its own flying experience and other particles flying experience.

Initially, the PSO algorithm randomly selects candidate solutions (particles) within the search space. During each iteration of the algorithm, each particle is evaluated by the objective function being optimized, determining the fitness of the solution. A new velocity value for each particle is calculated using the following equation:

$$v_i(t+1) = wv_i(t) + c_1 r_1 [\hat{x}_i(t) - x_i(t)] + c_2 r_2 [g(t) - x_i(t)] \quad (1)$$

The index of the particle is represented by i. So, $v_i(t)$ is the velocity of particle i at time t and $x_i(t)$ is the position of particle i at time t. Parameters w, c1, and c2 are user-supplied coefficients. The values r1 and r2 are random values regenerated for each velocity update. Value $\hat{x}_i(t)$ is the individual best candidate solution for particle i at time t, and g (t) is the swarm's global best candidate solution at time t. Once the velocity for each particle is calculated, each particle's position is updated by applying the new velocity to the particle's previous position using equation (2). This process is then repeated until some stopping condition is met. Figure 1 describes the flowchart of PSO algorithm

$$x_i(t+1) = x_i(t) + v_i(t+1) \quad (2)$$

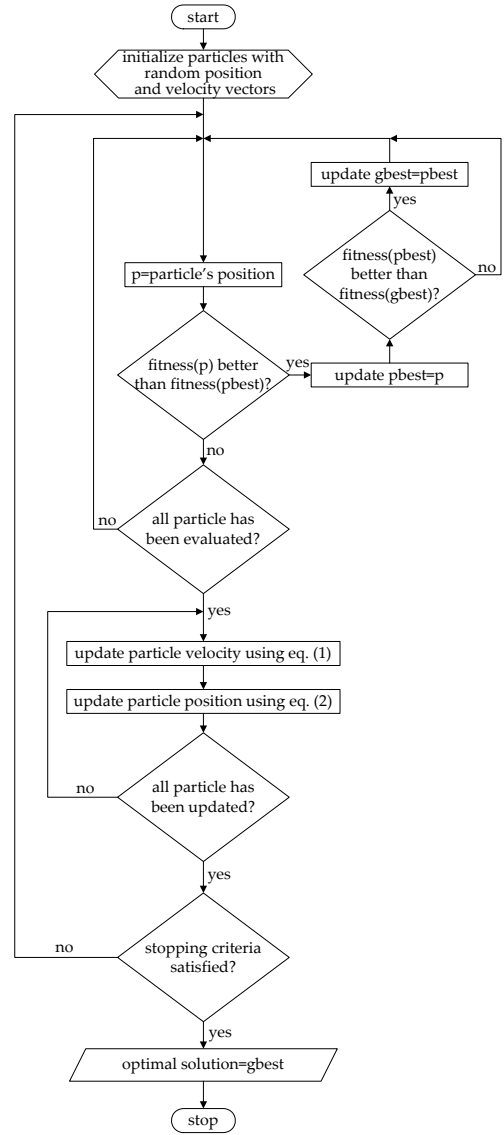

Fig. 1. Flowchart of particle swarm optimization algorithm.

## 4 TERM EXTRACTION USING PARTICLE SWARM OPTIMIZATION

In this section, we propose a new approach of term extraction, which takes into account several kinds of features, including domain relevance, domain consensus, term cohesion, first occurrence and length of noun phrase, to produce a list of terms.

Two steps are employed in our propose approach. First, terms are ranked to emphasize the most relevant from domain of input document; second, the score function is trained by the particle swarm optimization to obtain a suitable combination of feature weights.

### 4.1 Methodology
The goal of term extraction is to generate list of terms that are relevant to the domain of the input domain. Our proposed approach consists of the following steps:
1. Read the input document.
2. Preprocessing step consist of three sub tasks:
Syntactic parser does a syntactic analysis on every input sentence from input document, and produces a list of syntactic information (Noun Phrase-NP). Stop words should be filtered from each of the list of NP. Finally, the list of NP should be stemmed to produce list of clean NP, as the term candidate.
3. Each term candidate is associated with vector that contains five features.
4. The five features are used to calculate the term score and then rank the terms based on their score.

Our propose term extraction approach has two stages:
1. Training stages: This stage has the task to create a model for identifying terms using training documents. Features are extracted from training documents and used to train the swarm optimization model.
2. Extraction stages: This stage will choose terms from a test document (this document is different than that were used for training), using the model that has been made in the training stage.

Figure 2 shows our proposed term extraction model. Both stages choose a set of term candidate from their input documents, and then calculate the values of certain features for each candidate.

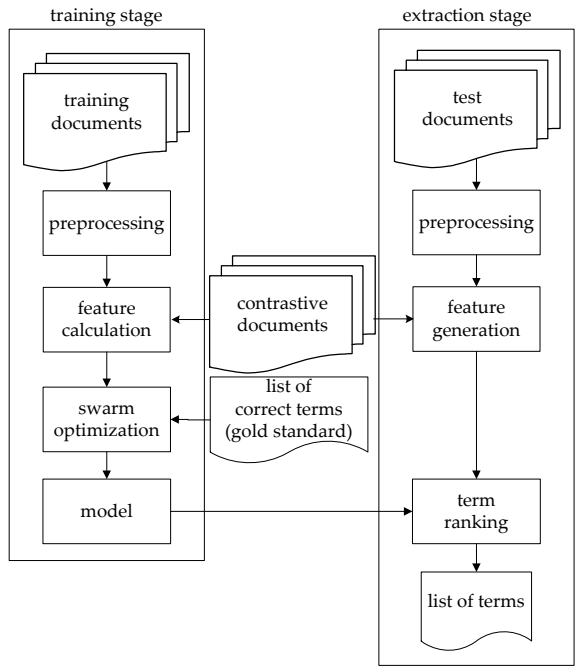

Fig. 2. The training and extraction stage processes

### 4.2 Feature Definition
In order to characterize the noun phrases in the documents we have adopted five features. These five features are calculated for each candidate term and used both in training and extraction stage. The features used are: domain relevance, domain consensus, term cohesion, first occurrence and length of noun phrase.

f1: *Domain relevance*-domain relevance can be given according to the amount of information captured in the target document with respect to contrastive documents. Let $D_i$ is the domain of interest (a set of relevant documents) and $\{D_1 ... D_n\}$ is sets of documents in another domain, domain relevance of a term t in class $D_i$ is computed as [12], [13]:

$$DR(t, D_k) = \frac{P(t|D_k)}{\max_{1 \le j \le n}(P(t|D_j))} \qquad (3)$$

where $(P(t|D_k))$ estimated as:

$$E(P(t|D_k)) = \frac{f_{t,k}}{\sum_{t' \in D_k} f_{t',k}} \qquad (4)$$

f2: *Domain consensus*-domain consensus measures the distributed use of a term in a domain $D_k$. Domain consensus is expressed as follows [12], [13]:

$$DC(t, D_k) = \sum_{d \in D_k} \left( P(t|d) \cdot \log_2\left(\frac{1}{P(t|d)}\right) \right) \qquad (5)$$

where:

$$E(P(t|d_j)) = \frac{f_{t,j}}{\sum_{d_j \in D_k} f_{t,j}} \qquad (6)$$

f3: *Term cohesion*-term cohesion is used to calculate the cohesion of the multi-word terms. The measure is proportional to the cooccurrence frequency and the length of the term [7], [8]:

$$TC(t) = \frac{|t| \cdot \log_{10}(f(t)) \cdot f(t)}{\sum_{w_i \in t} f(w_i)} \qquad (7)$$

f4: *First occurrence*-the main idea behind this feature is that important terms tend to occur at the beginning of documents. First occurrence is calculated as the number of words that precede its first appearance, divided by the number of words in the document. The resulting feature is a number between 0 and 1 representing the proportion of the documents before the term's first appearance [14].

f5: *Length of noun phrase*-candidate length is also a useful feature in extraction as well as in candidate selection, because the majority of terms are one or two words in length. Length of noun phrase score is calculated as its frequency times its length (in words) [20].

### 4.3 Term Generation
For a term t, a weighted score function, as shown in the following equation, is used to integrate all the feature scores mentioned in the previous section, where $w_i$ indicates the weight of $f_i$.

$$Score(t) = \sum_{i=1}^{5} w_i \cdot Score_{f_i}(t) \quad (8)$$

Moreover, the particle swarm optimization is used to obtain an appropriate set of feature weights. We have set the particle swarm optimization variables as follows: number of particles=40, maximum number of iterations=500, c1=2, c2=2 and w= (0.5 + (random/2)). During each iteration of the algorithm, each particle is evaluated using the fitness function as in (9). By applying particle swarm optimization, a suitable combination of feature weights could be found.

$$Fitness = \max\left(\sum_{i=1}^{|extracted|} |t_i \in goldstandard|\right) \quad (9)$$

where |extracted| is a number of terms extracted by the system and $|t_i \in goldstandard|$ is the number of terms that is a member of the gold standard (reference of correct terms).

### 4.4 Datasets
1. Quran (focus on verses about prayer): we use English translation to the meaning of the Quran (focus on verses about prayer) as the input document in the experiment. We separate the documents into a training documents and test documents.
2. Reuters-21578: the documents in the Reuters-21578 collection appeared on the Reuters newswire in 1987. In 1990, the documents were made available by Reuters for research purposes. We converted all the documents into 22 plain text file (reut2-000.txt until reut2-021.txt) and use it as contrastive documents.
3. Gold Standard: list of the Quran terms (focus on verses about prayer).

## 5 EXPERIMENTAL RESULTS

In the extraction stage, we evaluate the precision of our propose methods at 4 points: top 25, top 50, top 150 and top 250 terms using the following equation:

$$precision = \frac{\sum_{i=1}^{|extracted|} |t_i \in goldstandard|}{|extracted|} \quad (10)$$

We compare the terms extracted by the system with the gold standard that we have prepare before. Table 1 shows the term extraction precision for each feature for different number of terms evaluated.

TABLE 1
TERM EXTRACTION PRECISION FOR EACH FEATURE

| precision(feature) | number of terms | | | |
|---|---|---|---|---|
| | 25 | 50 | 150 | 250 |
| f1 | 0.800 | 0.820 | 0.607 | 0.552 |
| f2 | 0.880 | 0.760 | 0.673 | 0.596 |
| f3 | 0.880 | 0.780 | 0.673 | 0.596 |
| f4 | 0.800 | 0.740 | 0.650 | 0.610 |
| f5 | 0.880 | 0.740 | 0.600 | 0.584 |

We compare the precision of our propose method with four other known algorithms. The result show that our propose method based on particle swarm optimization can improve the precision of the extracted terms. Table 2 and Figure 3 show the comparison of the precision between swarm model and the four other algorithms (TFIDF, Weirdness, GlossaryExtraction and TermExtractor).

TABLE 2
COMPARISON OF THE TERM EXTRACTION PRECISION
(SWARM MODEL, TFIDF, WEIRDNESS,
GLOSSARYEXTRACTION AND TERMEXTRACTOR)

| precision(algorithm) | number of terms | | | |
|---|---|---|---|---|
| | 25 | 50 | 150 | 250 |
| TFIDF | 0.840 | 0.800 | 0.607 | 0.560 |
| Weirdness | 0.760 | 0.660 | 0.607 | 0.588 |
| GlossaryExtraction | 0.840 | 0.740 | 0.633 | 0.592 |
| TermExtractor | 0.840 | 0.800 | 0.647 | 0.564 |
| Swarm Model | 0.960 | 0.860 | 0.673 | 0.616 |

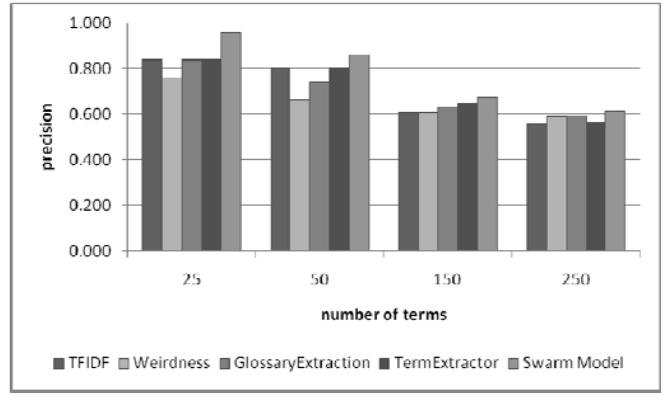

Fig. 3. Comparison of the term extraction precision (Swarm Model, TFIDF, Weirdness, GlossaryExtraction and TermExtractor)

## 6 CONCLUSION

We have presented a particle swarm optimization technique to improve term extraction precision. We choose five features to represent the term score: domain relevance, domain consensus, term cohesion, first occurrence and length of noun phrase. In the experiments, we use a translation of the meaning of the Quran (focus on verses of prayer) as an input document, both for training and testing phases. We separate the documents between training documents and test documents. Particles swarm optimization is trained using the training documents to determine the appropriate weight of each feature to produce the best score for each term. We conduct tests with the test document using the weight of each feature which is generated from the training stage to calculate the final score for each term to be extracted. Our experimental results show the use of particle swarm optimization technique can improve the precision of the extracted terms compared with four other known algorithms (TFIDF, Weirdness, GlossaryExtraction and TermExtractor).

## ACKNOWLEDGMENT

This project is sponsored by the Ministry of Science, Technology and Innovation under grant vote number: 79303.

**Mohammad Syafrullah** is a Ph.D candidate, Faculty of Computer Science and Information System in Universiti Teknologi Malaysia. He received his bachelor degree in Computer Science from Budi Luhur University, Indonesia in 1997. He received his master degree in Computer Science from Swiss German University, Indonesia in 2005. His current research interest includes Ontology Learning, Data Mining and Soft Computing.

**Dr. Naomie Salim** is an Associate Professor presently working as a Deputy Dean of Research & Postgraduate Studies in the Faculty of Computer Science and Information System in Universiti Teknologi Malaysia. She received her bachelor degree in Computer Science from Universiti Teknologi Malaysia in 1989. She received her master degree in Computer Science from University of West Michigan in 1992. In 2002, she received her Ph.D (Computational Informatics) from University of Sheffield, United Kingdom. Her current research interest includes Information Retrieval, Distributed Database and Chemoinformatic.